\documentclass[12pt,preprint]{aastex}
\usepackage{graphicx}
\usepackage{apjfonts}
\usepackage{natbib}

\slugcomment{Accepted for publication in ApJ.}

\shorttitle{Simultaneous observation of Mkn~501 with MAGIC and \textit{Suzaku} in 2006}
\shortauthors{Anderhub et al.}

\begin{document}


 \title{Simultaneous Multiwavelength observation of Mkn~501 in a low state in 2006}


%

\author{
H.~Anderhub\altaffilmark{1},
L.~A.~Antonelli\altaffilmark{2},
P.~Antoranz\altaffilmark{3},
M.~Backes\altaffilmark{4},
C.~Baixeras\altaffilmark{5},
S.~Balestra\altaffilmark{3},
J.~A.~Barrio\altaffilmark{4},
D.~Bastieri\altaffilmark{6},
J.~Becerra Gonz\'alez\altaffilmark{7},
J.~K.~Becker\altaffilmark{4}
% ...
\\
(The MAGIC collaboration)\\
and \\
 R.~Sato\altaffilmark{28}, 
 M.~Ushio\altaffilmark{28}, 
 J.~Kataoka\altaffilmark{29}, 
 G.~Madejski\altaffilmark{30}, 
 T.~Takahashi\altaffilmark{28}
}
\altaffiltext{1} {ETH Zurich, CH-8093 Switzerland}
\altaffiltext{2} {INAF National Institute for Astrophysics, I-00136 Rome, Italy}
\altaffiltext{3} {Universidad Complutense, E-28040 Madrid, Spain}
\altaffiltext{4} {Technische Universit\"at Dortmund, D-44221 Dortmund, Germany}
\altaffiltext{5} {Universitat Aut\`onoma de Barcelona, E-08193 Bellaterra, Spain}
\altaffiltext{6} {Universit\`a di Padova and INFN, I-35131 Padova, Italy}
\altaffiltext{7} {Inst. de Astrof\'{\i}sica de Canarias, E-38200 La Laguna, Tenerife, Spain}
% ...
\altaffiltext{28} {Institute of Space and Astronautical Science/JAXA, Kanagawa, 229-8510, Japan} 
\altaffiltext{29} {Research Institute for Science and Engineering, Waseda University, Tokyo, 169-8555, Japan}
\altaffiltext{30} {Kavli Institute for Particle Astrophysics and Cosmology, SLAC National Accelerator Laboratory, CA, USA}
% ...


\begin{abstract}
We present results of the multiwavelength campaign on the 
TeV blazar Mkn~501 performed in 2006 July, including MAGIC for 
the very-high-energy (VHE) $\gamma$-ray band and 
\textit{Suzaku} for the X-ray band.  A VHE $\gamma$-ray signal 
was clearly detected with an average flux above 200 GeV of 
$\sim$20\% of the Crab Nebula flux, which indicates a low state of source activity in this energy range.
No significant variability 
has been found during the campaign. The VHE $\gamma$-ray spectrum 
can be described by a simple power-law from 80 GeV to 2 TeV with a 
photon index of $2.8\pm0.1$, which corresponds to one of the 
steepest photon indices observed in this energy range so far for this object.  
The X-ray spectrum covers a wide range from 0.6 to 40 keV, 
and is well described by a broken power law, with 
photon indices of $2.257\pm0.004$ and $2.420\pm0.012$ 
below and above the break energy of $3.24^{+0.13}_{-0.12}$ keV. 
No apparent high-energy cut off is seen above the break energy. 
Although an increase of the flux of about 50 \% is observed in the X-ray band within the observation, the data indicate a consistently low state of activity for this source. Time-resolved spectra show an evidence for spectral hardening with a flux level.
A homogeneous one-zone synchrotron self-Compton (SSC) model can adequately
describe the spectral energy distribution (SED) from the X-ray 
to the VHE $\gamma$-ray bands with a magnetic field 
intensity $B=0.313$ G and a Doppler beaming factor $\delta = 20$, which are similar to the values in the past multiwavelength campaigns in high states. 
Based on our SSC parameters derived for the low state, we are able to 
reproduce the SED of the high state by just changing the Lorentz 
factor of the electrons corresponding to the break energy in 
the primary electron spectrum. This suggests that the variation 
of the injected electron population in the jet is responsible for 
the observed low-high state variation of the SED.
\end{abstract}



\section{Introduction}
Blazars are radio-loud active galactic nuclei (AGNs) viewed at small 
angles between the jet axis and our line of sight.  The bulk relativistic 
motion of the emitting plasma causes the radiation to be beamed in a forward 
direction, making the variability appear more rapid and the luminosity 
appear higher than in the rest frame due to the relativistic beaming 
effect~\cite[e.g.,][]{Ree66, Ghi93}.

Blazars with only weak or entirely absent emission lines in the optical band are classified 
as BL Lacertae Objects (BL Lacs). Their spectral energy distributions 
(SEDs: in $\nu F_{\nu}$) are characterized by a two-bump structure~\citep{Fos98}.  Since the 
discovery of the first extra-galactic TeV-photon emitter, 
Mkn421~\citep{Pun92}, very-high-energy (VHE: $E>$ 80 GeV) 
emission has been confirmed in more than 20 BL Lac objects.  The SEDs 
of many of those BL Lacs show the peaks of the lower energy bump at 
UV to X-ray energies. These objects belong to the sub-class known as 
"High-frequency peaked BL Lacs (HBLs)"~\citep{Pad95}.  Their non-thermal 
emission in this lower energy bump is commonly ascribed to synchrotron radiation from relativistic electrons, accelerated in the jet moving with 
relativistic bulk speed~\citep[e.g.,][]{Ghi98}. The two-bump structure 
in SEDs of HBLs has been well explained by synchrotron self-Compton (SSC) 
models, where the target photons of inverse-Compton scattering for the higher energy bump are the synchrotron photons produced by the same electron population~\citep[e.g.,][]{CandC02}.  In this model the 
high energy end of the electron spectrum is responsible for both X-ray and 
VHE $\gamma$-ray emission.  The observed correlations of the X-ray and VHE 
$\gamma$-ray fluxes during large flares of VHE $\gamma$-ray emitting 
HBLs~\cite[e.g.,][]{Tak96, Mar99, Kra01} provide strong experimental evidence 
for the SSC mechanism for HBLs. The target photons could also be produced in
the accretion disk~\citep{Der93} or in the broad-line region~\citep[e.g.,][]{Sik94}. 
Alternatively, the high-energy emission can be also due to pions produced by accelerated protons and ions and subsequent pion decay~\citep{Man93} or direct synchrotron emission from high-energy protons~\citep{Aha00}.

The results of applying emission models to the data can provide 
information on physical parameters of the jet, such as 
the co-moving magnetic field, the population of the accelerated 
electrons, the Doppler boosting factor and the size of the emitting 
region. HBLs often show strong flux variability on time scales of less than 1 hr~\citep{Gai96, Aha07, MAGIC501}. Hence, simultaneous multiwavelength 
observations over a wide energy range, covering in particular X-ray and VHE 
$\gamma$-ray bands, are essential to study the physics of these 
high-energy radiation emitters.

Until a few years ago, simultaneous multiwavelength observations 
were only possible during flaring states due to the low sensitivity 
of the participating $\gamma$-ray telescopes. In the VHE $\gamma$-ray 
band, new generation of Imaging Atmospheric Cherenkov Telescopes 
(IACTs) such as MAGIC, H.E.S.S., and VERITAS, can access the 
energy range from below 100\,GeV up to several TeV. These instruments also 
allow us to detect GeV-TeV $\gamma$-ray signals within short 
observation times (several hours) even in quiescent source states. 
A comparison of emission model parameters for several different source states may allow us to reveal the origin of the jet activity.

The \textit{Suzaku} X-ray satellite (see section~\ref{sec:Suzaku}) features the most sensitive 
instruments among current X-ray detectors for time-resolved 
coverage of a wide X-ray energy band, from the 
soft to the hard X-ray energies, well beyond 10\,keV. \textit{Suzaku} already performed observations of several 
known TeV HBLs and successfully obtained time-resolved spectra 
up to hard X-ray energies~\citep{Sato08, Tag08, Rei08}. The capability to perform successful multiwavelength observations for TeV-HBLs with MAGIC and \textit{Suzaku} even in quiescent states has been shown in \citet{Tag08} and \citet{Rei08}.

Mkn~501 (redshift $z = 0.034$) is categorized as an HBL and 
was the second established TeV blazar~\citep{Qui96}. In 1997 
this source went into a state of surprisingly high activity. 
The detected flux was 10 times higher than that of 
the Crab Nebula in the VHE $\gamma$-ray regime;  the 
high energy photons were observed up to $\sim$20 TeV~\citep{Aha01a}. 
At the same time, \textit{Beppo}SAX also observed a flaring activity 
in the X-ray band~\citep{Cat97, Pia98, Tav01}. The observed X-ray data 
showed an exceptionally hard spectrum with a synchrotron peak at 
(or above) $\sim$ 100\,keV. This represents a shift of at least 
two orders of magnitude with respect to previous observations. 
\citet{Gli06} organized a long-term monitoring campaign in 2004, also covering the X-ray and TeV energy bands. They confirm the presence 
of a direct correlation between X-ray and VHE $\gamma$-ray emission, 
which appears to be stronger when the source is brighter. In 2005, 
when MAGIC observed the object between May and July, 
the source flux varied by an order of magnitude. During the two 
most active nights, rapid VHE $\gamma$-ray flux variability with a doubling time of 
a few minutes was observed~\citep{MAGIC501}.  Several extensive 
SED studies based on multiwavelength observations of this object were reported~\citep[e.g.,][]{Kat99, Kra00, Sam00, Tav01, Ghi02}. However, the multiwavelength observations that included 
VHE $\gamma$-ray and X-ray instruments were only conducted during flaring states, but no simultaneous X-ray and VHE $\gamma$-ray data were available for a low state of activity.  

In 2006 July, a joint multiwavelength campaign between 
MAGIC, \textit{Suzaku}, and the Kungliga Vetenskaplika Academy (KVA) optical telescope\footnote{http://tur3.tur.iac.es/}
was organized to observe Mkn~501. We succeeded in obtaining clear detections 
from the simultaneous observations both in the VHE $\gamma$-ray band by 
MAGIC and in the X-ray band by \textit{Suzaku}. In this paper, we report 
the observational results of this campaign.  The observations and 
data reduction for both instruments are briefly described in Section 2. 
In Section 3, we present the results of the measured light curves and 
spectra for each energy band.  In Section 4, we discuss the application 
of a simple one-zone SSC model to the SEDs obtained in this campaign 
and compare those to the historical data from flaring states.  
Finally, we summarize our results in Section 5.
 
\section{Observations and data reduction}

\subsection{VHE $\gamma$-Ray band: MAGIC}

The MAGIC telescope 
is an IACT with a 17m diameter dish, located at the Canary Island 
La Palma (28$^{\circ}$.2~N, 17$^{\circ}.8$~W, 2225\,m\,a.s.l.).

In 2006, Mkn~501 was observed by MAGIC between July and September. 
The observations were performed in the so-called wobble mode~\citep{Dau97}, 
where the object is observed with an $0.4^{\circ}$ offset from 
the camera center. With this observation mode, ON- and OFF-data samples can be extracted from the same observation run; in our case, we used three OFF regions to estimate the background.
As a part of the multiwavelength campaign, intensive observations were conducted during three 
nights on July 18th, 19th, and 20th with a total observation 
time of 10.5 hr. In eight additional nights, MAGIC pointed at this source for only a few tens of minutes each time as a part of an extended monitoring program. In total, 4.2 hr were spent for these additional monitoring observations.  
After rejecting the data with anomalous trigger rate due to bad 
observation conditions, the remaining good quality data were further analyzed.
Data taken under large zenith angles ($> 35^{\circ}$) were also excluded to maintain the low energy threshold.

A shower image cleaning was applied based on the charge amplitude and time information 
in each pixel~\citep{MAGICCrab}.  Every cleaned event was 
parameterized using the so-called Hillas parameterization~\citep{Hillas}.
These parameters were used for $\gamma$/hadron separation 
and energy estimation of $\gamma$-ray events by means of the 
"Random Forest (RF)" method~\citep{MAGICRF}. In the RF method, 
Monte-Carlo simulated $\gamma$-ray samples~\citep{Maj05} with 
the same zenith angle range as the data were used as $\gamma$-ray 
training sample while real data were used as hadron-event background sample. The $\gamma$-ray signal was extracted on the basis 
of the DISP method~\citep{MAGICDISP}; a cut on the $\theta^{2}$ 
parameter (the squared angular distance between the nominal source 
position and the reconstructed $\gamma$-ray direction) was applied to determine the $\gamma$-ray signal~\citep{theta2}. Final spectra were 
derived using an unfolding technique~\citep{MAGICUnfold}. More 
detailed information on the standard analysis steps and performance 
of the MAGIC telescope are given in~\citet{MAGICCrab}. As quoted in that paper, we estimate a systematic energy scale error of 16\%, a systematic error of 11\% on the flux normalization (without the energy scale error), and a systematic slope error of $\pm0.2$.

For the multiwavelength campaign, based on 9.1 hrs of good quality data 
an excess of 1513 events over 26112 normalized background events yielding a significance of 8.0 $\sigma$ was obtained for the following analysis. We note that tighter cuts that only selected data of a shower image size > 350 photoelectrons (corresponding to a $\gamma$-ray energy peak of about 250 GeV) with a $\theta^{2} <  0.03\,{\rm deg}^{2}$ resulted in an increased 13.4 $\sigma$ significance.

\subsection{X-Ray band: \textit{Suzaku}}
\label{sec:Suzaku}

The joint Japanese-US satellite \textit{Suzaku}~\citep{Mit07}, 
launched successfully into orbit on 2005 July 10, covers a wide energy range of 0.2-600 keV.  \textit{Suzaku} carries four sets of 
X-ray telescopes~\citep{Ser07} each with a focal plane X-ray CCD 
camera~\citep[X-ray Imaging Spectrometer (XIS);][]{Koy07}, 
covering an energy range of 0.3-12 keV. Three of the XIS (XIS 0,2,3) 
detectors have front-illuminated (FI) CCDs, while the XIS-1 
utilizes a back-illuminated (BI) CCD. The merit of the BI CCD 
is its improved sensitivity in the soft X-ray energy band below 1 keV. 
\textit{Suzaku} also features a non-imaging collimated Hard 
X-ray Detector~\citep[HXD;][]{Tak07}, consisting of PIN silicon 
diodes for the lower energy band (10--70~keV) and GSO scintillators 
for the higher energy band (40--600~keV).  \textit{Suzaku} observations 
can be conducted using two 
default pointing positions, the XIS nominal position and the HXD 
nominal position. In this observation for the multiwavelength campaign, we used the HXD nominal 
position to maximize the effective area of the HXD. In the following 
analysis, the HXD/GSO data are not used because there is no significant 
detection above the 3$\sigma$ level.

X-ray observations by \textit{Suzaku} were carried out between 2006 July 18, 
18:33:00 UTC and 2006 July 19, 17:27:00 UTC (sequence 
number 071727010).  All XIS sensors were operated with 1/4 window 
option in order to reduce possible pile-up effects.  
In total 35 ks of good time intervals (GTIs) is obtained for each XIS and HXD 
detector after screening criteria as described in the following.


The XIS data used in this paper were reduced via the \textit{Suzaku} software version 2.0. 
The screening was based on the following criteria: 
(1) ASCA-grade 0,2,3,4 and 6 events were accumulated, and the CLEANSIS script 
was used to remove hot or flickering pixels.
(2) The time interval after the passage of the South Atlantic Anomaly is 
greater than 500 s.
(3) Data were selected to be 5$^{\rm o}$ in elevation above the rim of the Earth (ELV)
(20$^{\rm o}$ above the day-Earth rim). 
The XIS events were extracted from a circular region with a radius of $4.2'$ centered 
on the source peak, whereas the background was accumulated in an annulus with inner and 
outer radii $5.6'$ and $11.1'$, respectively. The response (RMF) and auxiliary files (ARF) are 
produced using the analysis tools {\tt XISRMFGEN} and {\tt XISSIMARFGEN} 
developed by the \textit{Suzaku} team~\citep{Ishi07}, which are included in the 
software package HEAsoft version 6.5. 


The HXD/PIN data (ver.2.0) were processed with basically the same 
screening criteria as those for the XIS, except for ELV$>5^{\rm o}$  
through night and day and a cutoff rigidity $>$8 GeV/$c$. The HXD/PIN 
instrumental background spectra were provided by the HXD team for each 
observation~\citep{Fuk06, Kak07}.  The HXD/PIN data also include 
another background component, the so-called, cosmic X-ray background (CXB).
In our analysis, we use the CXB spectrum~\citep{Gru99} as

\begin{equation}
 \frac{{\rm d}F}{{\rm d}E} =  9.0 \times 10^{-9} \left( \frac{E}{3 {\rm keV}} \right) ^{-0.29} \exp\left( \frac{-E}{40 {\rm keV}} \right) \frac{{\rm erg}} {{\rm cm}^{2}\ {\rm s}\ {\rm sr}\ {\rm keV}}
\end{equation}
The observed spectrum was derived assuming the PIN-detector response 
is isotropic for diffuse emission. Both the source and background 
spectra were made with identical GTIs, and the exposure was corrected 
for detector deadtime of 6.0\%. 
We use the response file version {\tt ae\_hxd\_pinhxdnom2\_20080129.rsp}.

Spectral analysis in the X-ray band was performed using XSPEC version 11.3.2.
Each XIS spectrum is binned such that each bin contains at least 40 counts. 
After the binning, we ignored bins with energies below 0.6 keV and above 10 keV. We also 
excluded bins between 1.7 and 1.9 keV because there exist 
large systematic uncertainties in the response matrices.
Based on a contemporaneous fit to the Crab spectra, \cite{Ser07} reported that the normalizations among the CCD sensors are slightly different (by a few \%).
A relative normalization of HXD/PIN detector to the CCD cameras also needs to be taken into account~\citep{Kak07}.
Therefore, the XIS-2, XIS-3 and HXD/PIN spectra were scaled by a constant factor with respect to the XIS-0 spectrum.


\section{Results}
\subsection{Light curves}
\begin{figure}[htbp]
 \centering
 \includegraphics[width=12cm]{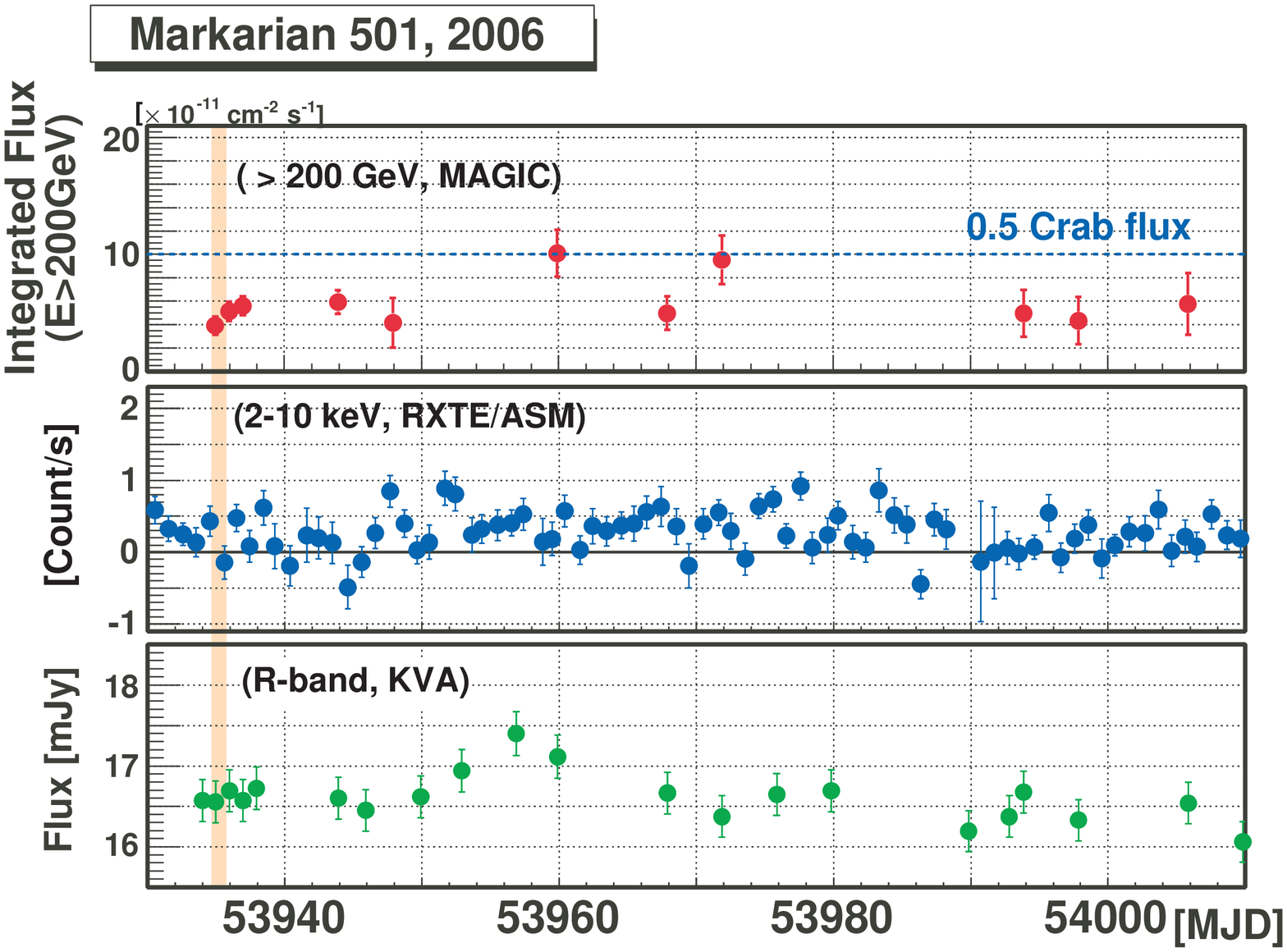}
   \caption{Diurnal multiwavelength light curves during the 
MAGIC observations of Mkn~501 in 2006 July-September. 
The vertical band represents the window of the \textit{Suzaku} 
pointing. \textbf{[Top]}: VHE $\gamma$-ray flux above 200 GeV as 
measured by MAGIC. The horizontal dotted line represents the half flux level of the Crab Nebula in this energy range. 
\textbf{[Middle]}: averaged daily X-ray count from RXTE/ASM. \textbf{[Bottom]}: optical \textit{R}-band flux by KVA.
}
   \label{Mkn5012006LC}
   \end{figure}

Figure~\ref{Mkn5012006LC} shows the diurnal light curves for Mkn~501 
in 2006 with the VHE $\gamma$-ray emission above 200 GeV as 
measured by MAGIC, public X-ray data taken by 
\textit{RXTE}/ASM\footnote{http://xte.mit.edu/ASM\_lc.html} 
and optical $R$-band data provided by the Tuorla Observatory 
Blazar Monitoring Program\footnote{more information at 
http://users.utu.fi/kani/1m/} using KVA.  In 2006, the source 
generally showed a low state of activity in the VHE $\gamma$-ray 
band, in contrast to 2005 when the flux varied by an order of 
magnitude up to 3.5 Crab units~\citep{MAGIC501}. Such a strong 
flare activity is not found neither in the X-ray nor the 
optical bands between 2006 July and September.

\begin{figure}[htbp]
  \centering
 \includegraphics[width=12cm]{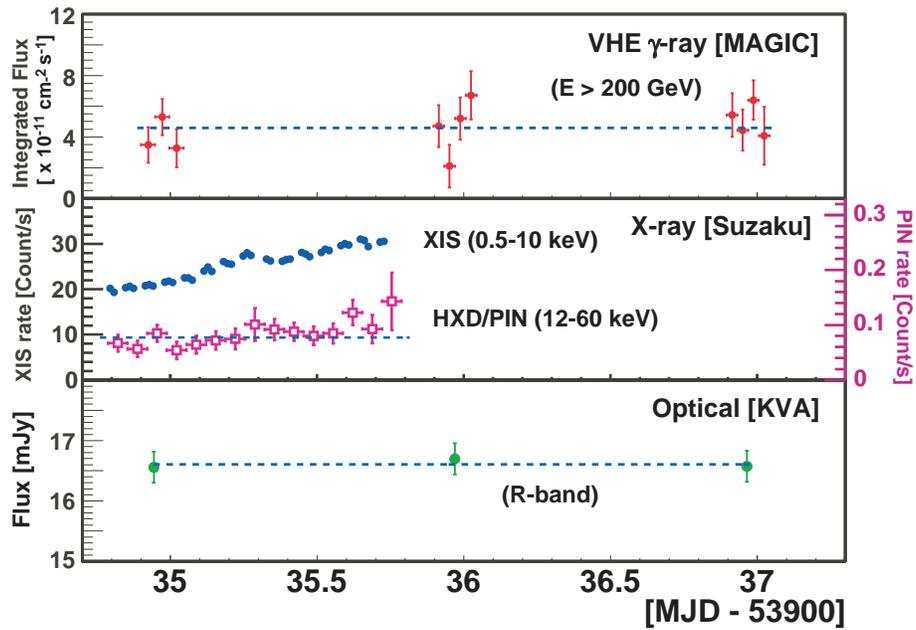}
   \caption{Light curves in different energy bands during this 
campaign. Each dotted horizontal line represents the average 
flux for each measurement. 
\textbf{[Top]}: VHE $\gamma$-ray flux measured by the MAGIC telescope. 
\textbf{[Middle]}: X-ray count rates measured by \textit{Suzaku} 
with the four XIS detectors (filled circle) and the 
HXD/PIN detector (open square).  
\textbf{[Bottom]}: optical \textit{R}-band flux measured by KVA.}
   \label{MWLC}
   \end{figure}

The light curves during the multiwavelength campaign in different 
energy bands taken by MAGIC, \textit{Suzaku} and KVA 
are shown in Figure~\ref{MWLC}.  The binning of the VHE $\gamma$-ray 
data is 1 hr.  The average integrated flux above 200 GeV is 
$(4.6\pm0.4)\times 10^{-11}\ {\rm cm}^{-2}\ {\rm s}^{-1}$ 
$(\chi^{2}/\rm{d.o.f.} = 10.1/10)$, which corresponds to about 
23\% of the Crab Nebula flux as measured by the MAGIC 
telescope~\citep{MAGIC501}. No significant variability 
is found. However, due to the low source flux level, 
we could only have seen variability if the flux were 
to increase by a factor of 2-3.


X-ray count rates of Mkn~501 recorded with the 4 XIS 
detectors (0.5-10 keV) and the HXD/PIN detector (12-60 keV) 
are plotted in the middle panel of Figure~\ref{MWLC}. 
Each point represents the rate in a time interval of 
1440 s for the XIS and 5760 s interval for the HXD/PIN. 
The X-ray count rate in the XIS shows clear variability ($\chi^{2}/\rm{d.o.f.} = 2\times10^4/4$1 for a constant flux fit). 
It gradually rises during the observation 
and an overall increase of about 50\% can be seen. 
The HXD/PIN count rate also seems to follow that 
increasing trend, but a fit with a constant value 
yields  $\chi^{2}/\rm{d.o.f.} = 13.3/14$.
The optical $R$-band flux is also consistent with 
a constant value with an average of 16.6$\pm$0.1 mJy.

\subsection{Spectra}

\begin{figure}[htbp]
  \centering
 \includegraphics[width=12cm]{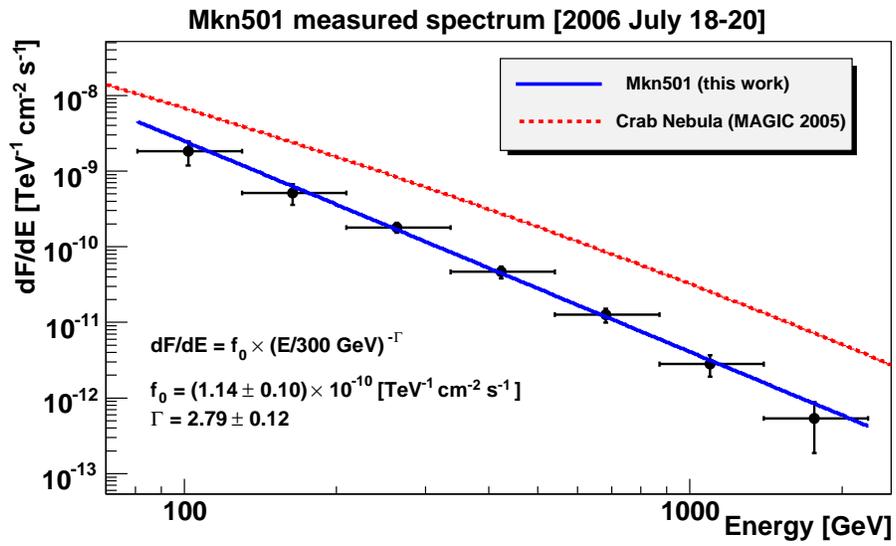}
   \caption{Measured differential energy spectrum at 
    VHE $\gamma$-rays for Mkn~501 with the MAGIC 
     telescope. The blue line corresponds to a simple power law fit. The fit parameters are listed in 
     the figure. For comparison, the measured MAGIC 
     Crab spectrum~\citep{MAGICCrab} is shown as red dashed line. Vertical bars denote the 1 
      $\sigma$ statistical error. Horizontal bars represent the size of the energy bins.
   }
   \label{Mkn501VHE_SP}
   \end{figure}

Figure~\ref{Mkn501VHE_SP} shows the unfolded differential 
spectrum in the VHE $\gamma$-ray band, averaged over 
the three days during the campaign. It is well described 
by a simple power law from 80 GeV to 2 TeV (with 
$\chi^2/{\rm d.o.f.}=1.85/5$):
\begin{equation}
 \frac{{\rm d}F}{{\rm d}E} =  (1.14 \pm 0.10) \times 10^{-10} \left( \frac{E}{0.3\ {\rm TeV}}\right)^{-2.79\pm0.12} \frac{\rm photons}{{\rm TeV}\ {\rm cm}^2\ {\rm s}}.
\end{equation}
An independent analysis gives results in very good agreement 
with those quoted numbers.  The flux level and the photon 
index of this spectrum are similar to those 
in the lowest state among the MAGIC results in 2005 which 
were presented on a night-by-night basis in~\citet{MAGIC501}.

\begin{figure}[htbp]
  \centering
 \includegraphics[width=15cm]{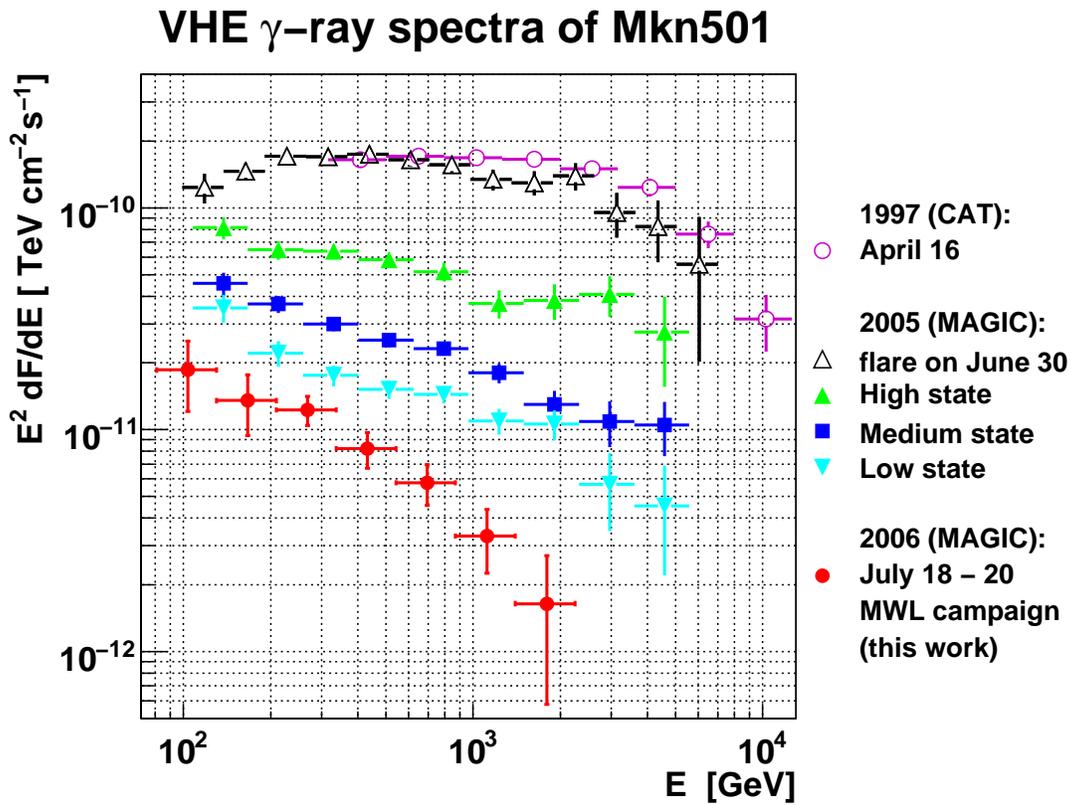}
   \caption[Measured VHE $\gamma$-ray spectra of Mkn~501 in 
      different activity states.]{Measured VHE $\gamma$-ray spectra 
      of Mkn~501 in different activity states. The CAT data were 
      taken from~\cite{Dja99}, the 2005 MAGIC data from~\cite{MAGIC501}, 
      the 2006 MAGIC data from this work. Vertical bars denote the 1 
      $\sigma$ statistical error. Horizontal bars represent the size of the energy bins.
   }
   \label{Mkn501VHEsp_all}
   \end{figure}

Figure~\ref{Mkn501VHEsp_all} shows the Mkn~501 $\gamma$-ray spectrum obtained during this multiwavelength 
campaign together with four spectra measured by MAGIC in 
2005~\citep{MAGIC501} and a spectrum by CAT Cherenkov imaging Telescope on 1997 April 16~\citep{Dja99}. As described in~\citet{MAGIC501}, 
the 2005 MAGIC data were subdivided into three groups, 
i.e., low-, medium- and high-states depending on the diurnal integral flux level. The other 
MAGIC spectrum corresponds to data taken during a strong flare on 2005 June 30. 
The CAT data were taken in 1997 during a previous multiwavelength
campaign with the \textit{BeppoSAX} X-ray satellite when 
the source was in a flaring state. Historically, spectra of 
Mkn~501 in the VHE $\gamma$-ray band have shown strong 
variability, and different features of the variability can be seen 
depending on the energy bands as shown in 
Figure~\ref{Mkn501VHEsp_all}. The difference in flux at 
$\sim$ 1 TeV reaches almost two orders of magnitude, 
while a difference of only a factor of 2-3 in the flux can be 
seen around 100 GeV. The spectrum gets harder as the source flux increases. The energy of 
the spectral peak, which we consider to be the inverse-Compton peak, can be seen above 100 GeV in the highest flux states (on 1997 April 16 by CAT and on 2005 June 30 by MAGIC), 
while the peak seems to be located well below 100 GeV in the 2006 observations.

\begin{figure}[htbp]
  \centering
 \includegraphics[width=9cm]{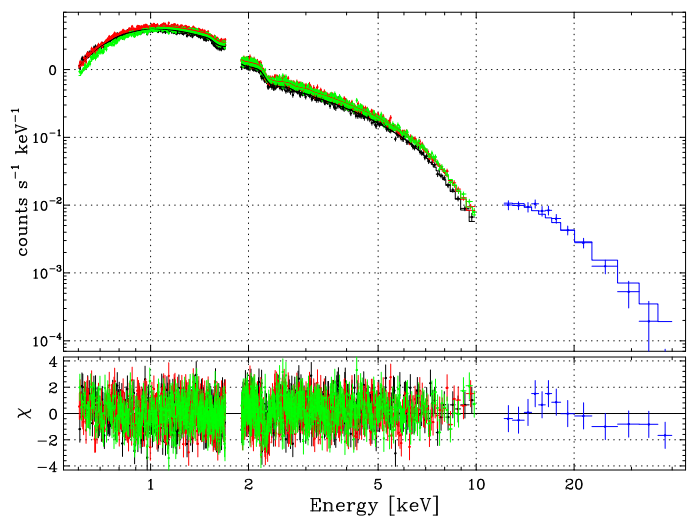}
   \caption{\textit{Suzaku} (XIS-0 [black], XIS-2[red], 
    XIS-3[green] and HXD/PIN [blue]) averaged spectrum of Mkn~501. The model
    plotted with the data is a broken power law obtained by a joint 
    fitting to these three XISs and HXD/PIN data. The parameters
    are shown in Table~\ref{Suzakufit}. The lower panels show the residuals for this broken power-law model. Vertical bars denote the 1 
      $\sigma$ statistical error.}
   \label{Suzakusp}
   \end{figure}

In the 0.6--40 keV X-ray band, we perform a joint fitting with three FI-XISs (XIS-0, 2 and 3: 0.6--10 keV) and HXD/PIN (12--40 keV) to derive a spectrum.
A fixed value of the galactic column density at $1.73 \times 10^{20}$ cm$^{-2}$~\citep{Sta92} is used for the galactic absorption.
The normalization of the HXD/PIN with respect to XIS-0 is fixed to be 1.15 as reported in~\citet{Kak07}.
A simple power-law fit 
yields a photon index of $\Gamma=2.292\pm0.002$, but this 
model gives an unacceptable fit (a reduced 
$\chi^2_\nu$=1.107 with 3893 d.o.f: probability$=2.4\times10^{-4}$\%).  
Fitting with a broken power law, 
we obtain a significantly improved fit (a reduced $\chi^2_\nu$ = 1.050 with 3891 d.o.f: probability$=1.5$\%) with the best-fit photon indices of $\Gamma_1 = 2.257\pm0.004$ 
and $\Gamma_2 =2.420\pm0.012$ below and above the 
break energy $E_{brk} = 3.24^{+0.13}_{-0.12}$ keV, respectively.
The constant factors for XIS-2 and XIS-3 are $1.033\pm0.003$ and $0.981\pm0.003$, respectively, which are within the values previously reported in \citet{Ser07}.
A double-broken power-law model does not improve the fit at all.
Figure~\ref{Suzakusp} shows background-subtracted folded 
count spectra of the three FI XISs and HXD/PIN with residuals 
for the broken power law, using the fixed Galactic column density.
The best-fit parameters and associated errors are summarized in Table~\ref{Suzakufit}.
These derived fit parameters are in good agreement with the results from the XIS data (0.6 -10 keV) fit alone. It suggests that there is 
no significant change (e.g., high-energy cut-off) in the spectrum 
between $E_{brk}$ and 40 keV. The photon index below the break 
($\Gamma_1$) clearly shows a softer value than 2. This indicates 
that the synchrotron peak is located at energies below the XIS range, i.e., lower than 0.6 keV.

\begin{figure}[htbp]
  \centering
 \includegraphics[height=9cm, angle=270]{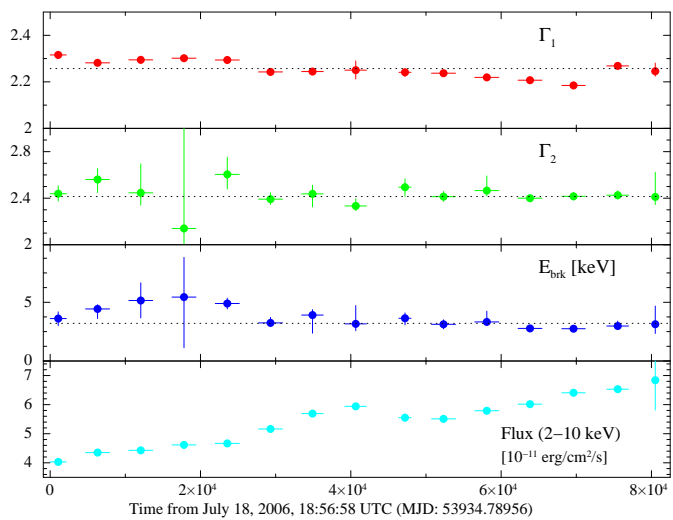}
   \caption{Temporal behavior of the fitting parameters to 
    the XIS data with a broken power-law model. Each point 
    represents a 5760s interval.  Photon indices below ($\Gamma_1$) 
    and above ($\Gamma_2$) the break energy ($E_{\rm brk}$) and model 
    flux between 2 and 10 keV are described.  A horizontal 
    dotted line in each panel represents an average value of each parameter. Note that because of a correlation between $\Gamma_2$ and $E_{\rm brk}$, the errors of $\Gamma_2$ can also vary depending on the uncertainties of $E_{\rm brk}$.}
   \label{Suzaku_modelLC}
   \end{figure}

\begin{figure}[htbp]
  \centering
 \includegraphics[height=7cm]{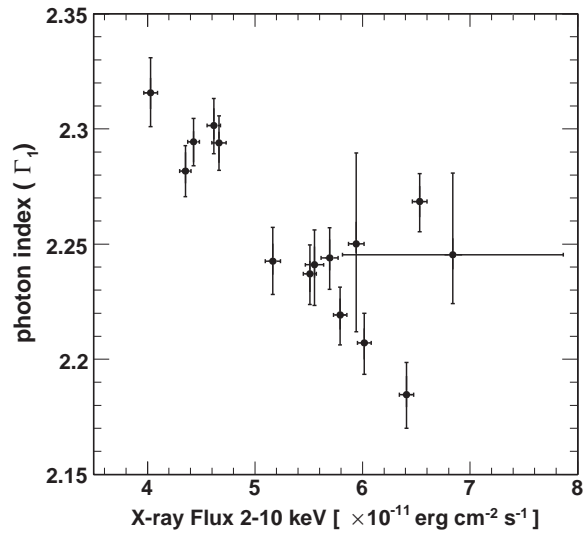}
   \caption{Scatter plot of X-ray flux between 2 and 10 keV and photon index below the break energy ($\Gamma_1$), which are taken from Figure~\ref{Suzaku_modelLC}.}
   \label{ind_flux}
   \end{figure}

We also attempted to fit time-resolved broken power-law spectra. The temporal behavior of three parameters, $\Gamma_1$,  
$\Gamma_2$, and the break energy, as well as the model flux between 2 and 10 keV, are presented in Figure~\ref{Suzaku_modelLC} with a 5760\,s interval. 
The normalization factors for the XISs were the same as stated above (see Table~\ref{Suzakufit}).
The model flux increases, similar to the XIS count rate. Figure~\ref{ind_flux} shows a scatter plot between the model flux between 2 and 10 keV and the photon index $\Gamma_1$.
A spectral hardening trend by $\sim$0.15 can be seen in $\Gamma_1$ as the flux increases with a correlation coefficient of $r=-(0.75^{+0.02}_{-0.20})$. On the other hand, the other parameters do not show strong, flux-dependent variations.

Although the X-ray data show some variability (increase by $\sim$ 50 \%), given the low flux level, 
it is still much less than the a factor of $\sim$ 2 that would be needed to 
clearly detect the variability in the VHE $\gamma$-ray range. 
Thus, in the following section we discuss the 
broad band SED of Mkn~501 using average spectra in both VHE 
$\gamma$-ray and X-ray bands taken during this multiwavelength campaign.


\section{Discussion}

\begin{figure}[htbp]
  \centering
 \includegraphics[width=10cm]{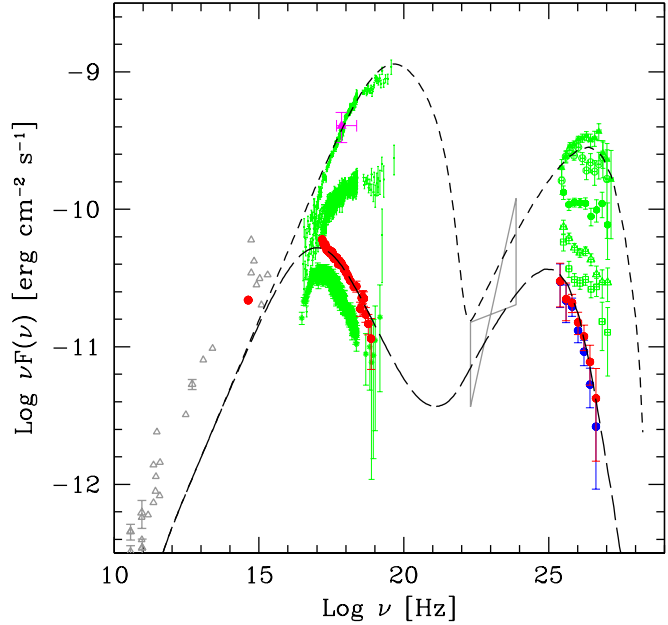}
   \caption[Overall SED of Mkn~501 as measured in 2006 July and historical 
data.]{Overall SED of Mkn~501 as measured in July 2006 and historical data. 
Red points represent energy fluxes from this campaign obtained by KVA (optical), 
\textit{Suzaku} (X-ray) and MAGIC (VHE $\gamma$-ray). VHE $\gamma$-ray 
fluxes are corrected by the "low-IR" EBL model of \cite{Kne04}. 
Corresponding measured VHE $\gamma$-ray fluxes are also plotted by blue points. 
Green points describe some historical X-ray and VHE $\gamma$-ray 
fluxes. The X-ray spectra were obtained by \textit{BeppoSAX} on 1997 
April 16 (the highest), 1997 April 29 (medium) and 1999 June (the 
lowest) (taken from~\citet{Tav01}). The X-ray flux measured by 
\textit{RXTE}/ASM on 2005 Jun 30 is also shown as pink point (see~\citet{MAGIC501}).
The historical VHE $\gamma$-ray spectra were obtained by MAGIC 
in 2005 taken from~\cite{MAGIC501}. They are also corrected for EBL absorption using the same Kneiske model. Grey points and a bow-tie 
are historical data taken from NASA Extragalactic database 
(radio-optical) and from \cite{Kat99} ($\gamma$-ray data observed 
by EGRET in 1996), respectively.  The long-dashed and short-dashed lines describe the SSC model 
based on \cite{Tav98, Tav01} for this campaign data and the "high" 
state, respectively. The model parameters can be seen in Table~\ref{Mkn501SSCp}.
}
   \label{Mkn501SSC}
   \end{figure}

Figure~\ref{Mkn501SSC} shows the overall SEDs of Mkn~501 with data 
obtained during this multiwavelength campaign as well as some 
historical data.  The flux in the VHE $\gamma$-ray band is 
corrected for absorption by the extra-galactic background light 
(EBL) using the "low-IR" model of \citet{Kne04}. In our optical data, 
the host galaxy contribution $(12.0\pm0.3)$ (mJy)~\citep{Nil07} 
has already been subtracted. 

Assuming a uniform injection of the electrons throughout a homogeneous 
emission region, we applied a one-zone SSC model, developed by \cite{Tav98, Tav01}, to our campaign data 
for estimating physical parameters of the emitting region. A spherical shape (blob) of radius $R$ 
is adopted for the emission region, filled with a 
tangled magnetic field of intensity $B$.  The electron distribution 
is described by a smoothed broken power-law energy distribution with 
slopes $n_1$ from $\gamma _{\rm min}$ to the break energy 
$\gamma _{\rm b}$ and $n_2$ up to a limit of $\gamma _{\rm max}$ 
and with a normalization factor $K$. The relativistic effect is 
taken into account by the Doppler beaming factor $\delta$.

The radius, $R$,  is selected to be $1.03\times 10^{15}$ cm, corresponding to the 
value reported in \cite{MAGIC501} for the SEDs observed in 2005.
$\gamma _{\rm max}$ is set to be $10^7$ since no cut-off in the high energy ends of both the X-ray and the VHE $\gamma$-ray spectra is detected.
$\gamma _{\rm min}$ is fixed at 1 as a nominal 
value because this parameter does not affect the emission in the energy bands of our data. Since the data do 
not clearly indicate the positions of the synchrotron and 
inverse-Compton peaks, we cannot fully constrain the SSC 
parameters by precise fits to the SEDs. In addition, variability 
timescales cannot be determined for the data because of the quiescent
state of the source.  Therefore, our aim is to reproduce the 
observed spectral behavior and correlations within a simple 
unifying picture rather than entering in details of the jet structure.

First, we applied the SSC model for the low state SED which 
is obtained during our multiwavelength campaign in 2006.
The one-zone SSC model can reproduce the measured X-ray and 
VHE $\gamma$-ray spectra in this low state of activity 
of the source as shown in Figure~\ref{Mkn501SSC}.  
However, it is apparent that the model in Figure~\ref{Mkn501SSC} underestimates the flux in a low energy range between radio and optical. 
Usually, homogeneous models cannot be used to explain the low frequency radio emission \citep[see e.g.,][]{Pia98} due to efficient self absorption. In previous studies, \citet{Kat01} used an inhomogeneous conical jet model proposed by~\citet{Ghi85} to explain radiation from the low radio frequency up to the ultraviolet of Mkn~501. Here, along the same lines, we consider the energy range between X-ray and TeV for our one-zone SSC model.

On the basis of model parameters for this low state, 
we attempted to reproduce the SED obtained during the 
flare on 2005 June 30, using the same SSC model.  
There are no simultaneous X-ray data other than 
\textit{RXTE}/ASM available at that time. The measured flux 
by the \textit{RXTE}/ASM (pink triangle in Figure~\ref{Mkn501SSC}) 
shows a compatible level in the X-ray spectrum to those 
taken by \textit{BeppoSAX} (green dots in Figure~\ref{Mkn501SSC}) 
on 1997 April 16. In addition, the VHE $\gamma$-ray spectrum 
taken by MAGIC on 2005 June 30, was almost equivalent 
to the spectrum measured by CAT on 1997 April 16, as 
shown in Figure~\ref{Mkn501VHEsp_all}. Therefore, we used 
this \textit{BeppoSAX} spectrum as a guide for the X-ray 
spectrum during the VHE $\gamma$-ray flare on 2005 June 30. 
With this, we can reproduce the SED in this high state just 
by changing $\gamma _{\rm b}$, the Lorentz factor of the 
electrons at the break energy in the electron spectrum.
The SSC models for the low and the high states of 
Mkn~501 are represented by dashed lines in Figure~\ref{Mkn501SSC}.
The derived parameters for these SSC models are listed 
in Table~\ref{Mkn501SSCp}.

In Table~\ref{Mkn501SSC_comp} we compare our results 
to some of the previous SED studies based on SSC models for Mkn~501.  
All of those were derived from applying one-zone SSC 
models to actual observational data. Not all studies 
used simultaneous X-ray and VHE $\gamma$-ray data.  
In fact, most simultaneous X-ray and VHE $\gamma$-ray 
data were taken during the huge outbursts in 1997.
Nevertheless, the SSC model parameters of $\delta$ and 
$B$ from our multiwavelength campaign in 2006 indicate 
values similar to those of previous works for different 
flux states, apart from models of \cite{Sam00} (lower $B$) 
for the medium flux state in 1998\footnote{however, they 
did not take into account the absorption by EBL in the 
VHE $\gamma$-ray data.}, and of \cite{Kon03} (higher 
$\delta$ and lower $B$) for the high flux state in 1997.
Those parameters are also consistent with values in~\citet{Bed99}, who constrain the parameters of the emission region based on the variability time scale during the 1997 April 15-16 flaring activity. Note again that only our work used simultaneous X-ray and 
VHE $\gamma$-ray data taken in a low flux state.
 
\cite{Tav01} attempted to model different emission 
states of Mkn~501 in 1997 and 1999 by mainly changing 
the break energy of electrons, slightly modifying its 
spectral slopes and number density, and by keeping 
other parameter unchanged. Also \cite{Pia98} could 
reproduce different flux states by changing only the 
electron distribution, keeping the same values for 
other parameters. In these frameworks, the electron 
spectrum is the key component representing the different 
activity states of Mkn~501; especially, $\gamma _{\rm break}$ 
can play a major role there.

From the physical point of view, in the context of the widely 
discussed diffusive shock acceleration models~\citep{Kir98, Hen99} 
the variations of $\gamma _{\rm break}$ (and the other parameters 
specifying the particle acceleration) could be explained by 
changes of the parameters determining the efficiency of the 
acceleration mechanisms (such as the parameters characterizing 
the turbulence). A deeper discussion of this point is clearly 
beyond the scope of this paper.

\section{Summary}

We present first results of a multiwavelength campaign for the 
TeV blazar Mkn~501 during its low state of activity observed by 
MAGIC and \textit{Suzaku} in 2006 July.

\begin{itemize}

\item VHE $\gamma$-ray signals were clearly detected at a 13.4 $\sigma$ level from 9.1 hrs of data taken during the campaign.
The average integrated flux above 200 GeV corresponds to 
about 23\% of the Crab Nebula flux without significant variability 
detected in the data.  The spectrum in the VHE $\gamma$-ray band 
is well described by a simple power law from 80 GeV to 2 TeV with 
a photon index of ${2.8\pm0.1}$.  The flux level and the photon 
index of this measured spectrum are compatible with those found 
in the lowest state among the MAGIC Mkn~501 observations in 2005. 

\item The X-ray spectral shape derived from the \textit{Suzaku} 
data from 0.6 keV up to 40 keV is well described by a broken power-law without a need for a 
high energy cut-off. The derived photon index suggests that the 
synchrotron peak is located below 0.6 keV. The flux level in X-ray 
showed a low state of activity, in similarity to the VHE $\gamma$-ray 
flux. As the flux shows a small increase ($\sim$50\,\%), the spectral index below the break energy shows a hardening trend by $\sim$0.15.

\item The overall SED in the low state during our multiwavelength 
campaign can be well represented by a homogeneous one-zone SSC model.  
Based on the SSC parameters for this low state, the high state SED in 
2005 can be reproduced by changing only the Lorentz factor of the 
electrons corresponding to the break energy of the electron spectrum.   This suggests 
that the variation of the injected electron population in the jet is 
responsible for the observed variation of the SED. In particular, 
the electron energy at the spectral break could be a key parameter 
to represent the different activity states of Mkn~501. 

\end{itemize}

\acknowledgements
We thank the Instituto de Astrofisica de Canarias 
for the excellent working conditions at the Observatorio del Roque 
de los Muchachos in La Palma. The support of the German BMBF and MPG, 
the Italian INFN and Spanish MCINN is gratefully acknowledged.  
This work was also supported by ETH Research Grant TH 34/043, 
by the Polish MNiSzW Grant N N203 390834, by the YIP of the Helmholtz Gemeinschaft and by the U.S. 
Department of Energy contract to SLAC no. DE-AC3-76SF00515.




\clearpage

\clearpage
   
%

\clearpage

\begin{deluxetable}{ccccccccc}
\tablecaption{Best-Fit parameters for \textit{Suzaku} data.\label{Suzakufit}}
\tabletypesize{\scriptsize}
\tablewidth{19cm}
\rotate
\centering                         
\startdata       
\hline \hline
Model & $\Gamma_{1} $ &    $E_{\rm brk}$ &  $\Gamma_{2} $    &  const. & const. & const. & $F_{2-10{\rm keV}}$ &  $\chi ^2_{\rm \nu}$/d.o.f. \\  
(1)  & (2) & (3)  & (4)  & [XIS2] (5) &  [XIS3] (6) & [HXD/PIN] (7) &  (8) & (9) \\
\hline \hline
\multicolumn{9}{c}{\textit{Suzaku} XIS + HXD/PIN} \\
\hline
 power law &   $2.292\pm0.002$ & -- &  -- & $1.033\pm0.003$ & $0.981\pm0.003$ & 1.15 (fixed) & $5.370\pm0.012$ & 1.107/3893 ($2.4\times10^{-4}$\%) \\
broken power law & $ 2.257\pm0.004$   & $3.24^{+0.13}_{-0.12}$ &  $2.420\pm0.012$  & $1.033\pm0.003$  & $0.981\pm0.003$ & 1.15 (fixed) & $5.311^{+0.014}_{-0.019}$  &   1.050/3891  (1.5\%) \\
\hline\hline
\multicolumn{9}{l}{Galactic column density: $N_H=1.73\times10^{20}\ [{\rm cm}^{-2}]$ (fixed)} \\
\enddata
\tablecomments{Col. (1): model used to fit the data. Col. (2): photon index for the power law model, or low-energy photon index for the broken power law model. Col. (3): break energy [keV] for the broken power law model. Col. (4): high-energy photon index for the broken power law model. Col. (5,6,7): constant factors with respect to XIS1, for XIS2, XIS3 and HXD/PIN, respectively. Col. (8): flux in the 2-10 keV band, in units of $10^{-11}$ [erg cm$^{-2}$\ s$^{-1}$]. Col. (9): reduced $\chi^2/$degrees of freedom and corresponding probability.}

\end{deluxetable}

\begin{deluxetable}{cccccccccc}   
\tablecaption{SSC model parameters of Mkn~501.\label{Mkn501SSCp}}
\centering                         
\startdata
\hline\hline                
data & $R$ & $\delta$ &  $\gamma_{\rm min}$ & $\gamma_{\rm br} $ & $ \gamma_{\rm max} $& $B$ & $K$ & $n_1$ & $n_2$ \\
	   & cm   & 				&					&				&		& Gauss & particle/${\rm cm}^{3}$ &  & \\
\hline 
2006 (low) & $1.03 \times 10^{15}$ & 20 & 1 &  $6\times 10^{4} $ & $1 \times 10^{7} $ & 0.313 & $8.5\times 10^{4} $& 2 & 3.9 \\
2005 (flare) & $1.03 \times 10^{15}$ & 20 & 1 & $1.3\times 10^{6} $ & $1 \times 10^{7} $& 0.313 & $8.5\times 10^{4} $& 2 & 3.9 \\
\hline
\enddata
\end{deluxetable}

\begin{deluxetable}{cccll}   
\tablecaption{Comparison of the SSC model parameters, $\delta$, $B$ and $R$ to previous studies for Mkn~501. \label{Mkn501SSC_comp}}
\centering                        
\startdata    
\hline\hline                
$\delta$ &  $B$ [G] &  $R$ [cm]  & Flux State\tablenotemark{1} & reference \\ 
\hline
15 & 0.8 &  $5 \times 10^{15}$ & H (1997), M(1997) and L &  \cite{Pia98} \\
15 & 0.2 & $4.5\times 10^{15}$ & L &  \cite{Kat99}\\
25 & 0.1 & $4\times10^{16}$ & H (1998) & \cite{Sam00} \\
25 & 0.03 & $2\times10^{15}$ & M (1998) & \cite{Sam00} \\
14 & 0.15 &  $2.9 \times 10^{15}$ & H (1997) &  \cite{Kat01}\\
 14 & 0.15 & $4.2 \times 10^{15}$ & M (1997) &  \cite{Kat01} \\
10 & 0.32 & $1.9 \times 10^{15}$ & H (1997), M (1997) and L &  \cite{Tav01} \\
11 & 0.2 & $10 \times 10^{15}$ &  L &  \cite{Kin02}\\
50 & 0.04 &  $3.5 \times 10^{15}$ &  H (1997) & \cite{Kon03} \\
20 & 0.313 &  $1.03 \times 10^{15}$ &  H (2005), L (2006) &  this work\\
\hline
\enddata
\tablenotetext{1}{H: high state, M: medium state, L: low state (among historical data in X-ray. \textit{Beppo}SAX (green) data in Figure~\ref{Mkn501SSC} represent those states). Numbers represent the year when the corresponding data were taken. Previous works for "L" use data taken in different years in different energy bands. }

\end{deluxetable}


\end{document}